\documentclass[referee]{chjaa}          

\usepackage{graphicx,times,natbib}          
\usepackage{epsfig}

\newcommand{\bd}{\begin{displaymath}}
\newcommand{\ed}{\end{displaymath}}

\begin{document}

\title{The advection-dominated accretion
flow+thin accretion disk model for two low-luminosity active
galactic nuclei: M81 and NGC4579 }

   \volnopage{Vol.0 (200x) No.0, 000--000}      
   \setcounter{page}{1}           

\author{Ya-Di Xu \inst{1}\mailto{}, Xinwu Cao \inst{2}}

 \institute{Physics Department, Shanghai Jiao Tong University,800
Dongchuan Road, Min Hang, Shanghai, 200240, China\\
    \email{ydxu@sjtu.edu.cn; cxw@shao.ac.cn}
  \and
     Shanghai Astronomical Observatory, Chinese academy of
Sciences,80 Nandan Road, Shanghai 200030, China\\
 }

   \date{Received~~2001 month day; accepted~~2001~~month day}

\abstract {
It was found that the advection-dominated accretion flow
(ADAF)$+$thin disk model calculations can reproduce the observed
spectral energy distributions (SED) of the two low-luminosity active
galactic nuclei (AGNs), provided they are accreting at $\sim
0.01-0.03$ Eddington rates and the thin disks are truncated to ADAFs
at $\sim 100R_{\rm S}$ ($R_{\rm S}$ is the Schwarzschild radius) for
M81 and NGC4579 \citep{qdnh99} . However, the black hole masses
adopted in their work are about one order of magnitude lower than
recent measurements on these two sources. Adopting the well
estimated black hole masses, our ADAF$+$thin disk model calculations
can reproduce the observed SEDs of these two low-luminosity active
galactic nuclei, if the black hole is accreting at $2.5\times
10^{-4}$ Eddington rates with the thin disk truncated at $R_{\rm
tr}=120R_{\rm S}$ for M81 ($\dot{m}=3.3\times10^{-3}$ and $R_{\rm
tr}=80R_{\rm S}$ are required for NGC4579).
The transition zones with temperature from the thin disk with $\sim
10^{4-5}$ to $\sim 10^{9-10}$~K in the ADAF will inevitably emit
thermal X-ray lines, which provides a useful diagnosis on their
physical properties. The observed widths of the thermal X-ray iron
lines at $\simeq 6.8$~keV are consistent with the Doppler broadening
by the Keplerian motion of the gases in the transition zones at
$\sim 100R_{\rm S}$. We use the structure of the transition zone
between the ADAF and the thin disk derived by assuming the turbulent
diffusive heat mechanism to calculate their thermal X-ray line
emission with standard software package Astrophysical Plasma
Emission Code (APEC). Comparing them with the equivalent widths of
the observed thermal X-ray iron lines in these two sources, we find
that the turbulent diffusive heat mechanism seems to be unable to
reproduce the observed thermal X-ray line emission. The test on the
evaporation model for the accretion mode transition with the
observed thermal X-ray line emission is briefly discussed.
}

   \authorrunning{The ADAF$+$thin accretion disk model}         
   \titlerunning{Xu Y.D. \& Cao X. }  

\maketitle
\section{Introduction}

It is well accepted that many astrophysical systems are powered by
black hole accretion. The standard thin disks (i.e., geometrically
thin and optically thick accretion disks) can successfully explain
most observational features of the black hole accretion systems
\citep{ss73}. The ultraviolet/optical continuum emission observed in
luminous quasars is usually attributed to the thermal radiation from
the standard disks (SD) surrounding the massive black holes in
quasars \citep*[e.g.,][]{sm89}. However, the standard disk model is
unable to reproduce the spectral energy distributions (SED) of many
sources (e.g., Sgr A$^*$) accreting at very low rates, and the
advection dominated accretion flows (ADAF) were suggested to be
present in these sources \citep{ny94,ny95}. In the ADAF model, most
released gravitational energy of the gases in the accretion flow is
converted to the internal energy of the gas, and the ADAF is hot,
geometrically thick, and optically thin. Only a small fraction of
the released energy in ADAFs is radiated away, and their radiation
efficiency is therefore significantly lower than that for standard
thin disks \citep*[see][for a review and references therein]{nmq98}.
The ADAF model can successfully reproduce the observed SEDs in many
black hole systems accreting at low rates
\citep*[e.g.,][]{lackny96,nmy96,nmg98}.

There is a critical accretion rate $\dot{m}_{\rm crit}$, above which
the ADAF is suppressed and a standard thin disk is present. The ADAF
can co-exist with the standard thin disk, when it is accreting at
rates slightly lower than the critical value $\dot{m}_{\rm crit}$.
In this case, the ADAF is present in the inner region near the black
hole and connects to a standard thin disk at a certain transition
radius $R_{\rm tr}$ \citep*[see][for a review and references
therein]{nmq98}. \citet{nmy96} proposed the ADAF$+$standard thin
disk systems to model the observed spectra of the black hole
accretion systems with moderate accretion rates. \citet{n96} found
that many different spectral states observed in black hole X-ray
binaries can be understood as a sequence of ADAF$+$thin disk models
with varying $\dot{m}$ and $r_{\rm tr}$ (where $r_{\rm tr}$ is the
outer radius of ADAF in units of Schwarzschild radius, $R_{\rm
S}=2GM_{\rm bh}/c^{2}$). This scenario was explored in more detail
by several authors \citep{emn97,encgz98}, and they found that the
different spectral states of X-ray binaries (i.e.,  the quiescent
state, low state, intermediate state or high state) correspond to
different accretion rates. It is believed that standard thin disks
(or slim disks) are present in luminous active galactic nuclei
(AGN), while ADAFs are present in those low-luminosity AGNs
accreting at relatively low rates. The ADAF$+$thin disk systems are
required for modeling on a variety of observations of AGNs accreting
at the moderate rates \citep*[e.g.,][]{qdnh99,cao03}.

The physical mechanisms of the transition from a thin disk to an
ADAF and its structure are still quite uncertain, though a few
different scenarios were suggested
\citep[e.g.,][]{mmh94,h96,lymmx99}. \citet{mmh94} proposed an
"evaporation" mechanism, in which the disk is evaporated as heated
by electron conduction from a hot corona, and then a quasi-spherical
hot accretion flow is formed. An alternative scenario was suggested
by \citet{h96}, in which a "turbulent diffusive heat" mechanism is
employed to explore the structure of the ADAF$+$thin disk system.
This model was further developed, and the global structure of the
ADAF$+$thin disk system was derived numerically
\citep{mk2000,mknn2000}. All of these model calculations show that a
rapid decrease of temperature occurs within a narrow zone between
the inner ADAF and the outer standard thin disk at $\sim r_{\rm
tr}$.


\citet{qdnh99} showed that the optical/UV to X-ray emission detected
from the nuclei of M81 and NGC 4579 can be well explained by an
optically thick, geometrically thin accretion disk extends down to
~100~$R_{\rm S}$ (inside of which an ADAF is present), provided
their accretion rates are $\dot{m}\sim 0.01$. The optical/UV bumps
observed in these two sources can be attributed to the thermal
emission from the outer standard thin disks, while their hard X-ray
emissions are dominantly from the inner ADAFs. In their model
calculations, the same central black hole mass $M_{\rm
bh}=4\times10^6$~$M_\odot$ is adopted for these two sources, which
are about an order of magnitude underestimated compared with the
recent estimates \citep{d03,b01}. The iron K$\alpha$ lines were
observed in these two sources \citep{i96,t98}. In addition to narrow
iron K$\alpha$ lines at 6.4 KeV, the broad line components centered
at $E_{{\rm K}\alpha}=6.79$~KeV were also observed in these two
sources \citep{dgds04}. The thermal X-ray line emission is a useful
diagnosis on black hole accretion
systems\citep*[e.g.,][]{nr99,xu06}. \citet{dgds04} found that the
observed lines from these two sources are too broad to be the
thermal line emission from the host galaxy, and they suggested that
these broad emission lines may probably be from the transition zones
between the ADAFs and the outer standard thin disks. The temperature
of the inner edge of the thin disk is around $10^4$~K, and therefore
the temperature of the transition zone extends from $\sim 10^4$ to
$\sim 10^{8-9}$~K while connecting to the ADAF. Thus, the thermal
X-ray line emission can naturally originate from such transition
zones. The observed widths of the broad lines are roughly consistent
with the Doppler broadening by the Keplerian motion of the gases in
the transition zones at $\sim100~R_{\rm S}$ \citep{qdnh99}.

In this work, we re-investigate the ADAF$+$thin disk systems in
these two low-luminosity active galactic nuclei, NGC 4579 and M81,
adopting the black hole masses estimated by \citet{d03,b01}.  The
thermal iron K$\alpha$ line emissions from the transition zones of
the ADAF to the thin disk in these two low-luminosity active
galactic nuclei, NGC 4579 and M81, are calculated with the
transition model based on the turbulent diffusive heat mechanism
developed by \citet{h96}.



\section{The ADAF+thin disk model}

The observed SEDs of two low-luminosity AGNs, M81 and NGC4579, were
reproduced with the ADAF$+$standard thin disk model by
\citet{qdnh99}. In this ADAF$+$standard thin disk model, no smooth
physical connection between the ADAF and the outer standard thin
disk at $r_{\rm tr}$ has been included. The resulted spectra are
simply the combination of the spectra from the ADAF and the outer
thin disk. \citet{qdnh99}'s model calculations showed that the best
fits to the observed SEDs of these two sources require the
transition radius $r_{\rm tr}\simeq 100$ for both sources. The
accretion rates are $\dot{m}=0.03$ and $\dot{m}=0.01$ for NGC4579
and M81, respectively. The viscosity parameter $\alpha=0.1$, the
ratio of gas to total pressure $\beta=0.9$, the fraction of the
released energy directly heating the electrons $\delta=0.01$, are
adopted in their calculations. In their model calculations, the same
central black hole mass $M_{\rm bh}=4\times10^6$~$M_\odot$ is
adopted for these two sources, which are about an order of magnitude
underestimated. The masses of these two black holes were estimated
as $M_{\rm bh}\simeq 7\times10^7$~$M_\odot$ for M81, and $M_{\rm
bh}\sim 5\times10^7$~$M_\odot$ for NGC4579\citep{d03,b01}. Thus, we
have to re-calculate the global structures of the ADAFs surrounding
the black holes in these two sources. { We employ the approach
suggested by \citet{m00} to calculate the global structure of an
accretion flow surrounding a Schwarzschild black hole in general
relativistic frame. All the radiation processes are included in the
calculations of the global accretion flow structure \citep*[see][for
the details and the references therein]{m00}.} It was pointed out
that a significant fraction of the viscously dissipated energy could
go into electrons by magnetic reconnection, if the magnetic fields
in the flow are strong \citep{bl97,bl00}. They argued that $\delta$
can be as high as $\sim 0.5$. Therefore, we adopt a conventional
value of $\delta=0.3$ \citep*[e.g.,][]{wyc07}, and tuning the
accretion rates $\dot{m}$ to fit their observed optical/UV/X-ray
continuum spectra. The irradiation of the outer thin disk by the
ADAF is included in our calculations. We include an empirical color
correction given by \citet{c02} for the thermal emission from the
outer thin disk, which can reproduce the non-LTE disk spectral model
of \citet{h01}.

{ The flux due to viscous dissipation in the outer region of the
disk is \begin{equation}F_{\rm vis}(R)\simeq {\frac {3GM_{\rm
bh}\dot M}{8\pi R^3}}, \label{fvis} \end{equation}which is a good
approximation for $R_{\rm tr}\gg R_{\rm in}$. The outer thin disk
may be irradiated by the incident photons from the inner ADAF
region. The flux due to the irradiation in the outer thin disk
$F_{\rm irr}(R)$ can be calculated by assuming the incident photons
are reprocessed as thermal radiation from the thin disk, while the
global structure of the inner ADAF is available \citep*[see][for the
details]{cw06}. The local disk temperature of the thin cold disk is
\begin{equation}T_{\rm disk}(R)={\frac {[F_{\rm vis}(R)+F_{\rm
irr}(R)]^{1/4}}{\sigma_{\rm B}^{1/4}}}, \label{tdisk}
\end{equation}by assuming local blackbody emission. In order to calculate the
disk spectrum, we include an empirical color correction for the disk
thermal emission as a function of radius. The correction has the
form \citep{c02} \begin{equation}f_{\rm col}(T_{\rm disk}) =
f_\infty - \frac{(f_\infty - 1) [1 +
                     \exp(-\nu_{\rm b}/\Delta\nu)]} { 1 +
                     \exp[(\nu_{\rm p} -\nu_{\rm b})/\Delta\nu]}, \label{fcol}
\end{equation}where $\nu_p \equiv 2.82k_B T_{\rm disk}/h$ is the peak
frequency of a blackbody with temperature $T_{\rm disk}$.  This
expression for $f_{\rm col}$ goes from unity at low temperatures to
$f_\infty$ at high temperatures with a transition at $\nu_{\rm b}
\approx \nu_{\rm p}$. \citet{c02} found  that $f_\infty = 2.3$ and
$\nu_b = \Delta\nu = 5\times 10^{15}$\,Hz do a reasonable job of
reproducing the model disk spectra of \citet{h01}. The disk spectra
can therefore be calculated by \begin{equation}L_\nu =8\pi^2 \left(
{\frac {GM}{c^2}} \right)^2 {\frac{h \nu^3}{c^2} }
     \int\limits_{r_{\rm tr}}^\infty
     {\frac{r dr}{ f_{\rm col}^4[\exp(h\nu/f_{\rm col} k_B T_{\rm disk}) - 1]}}.
\end{equation} }

\begin{figure}
\plotone{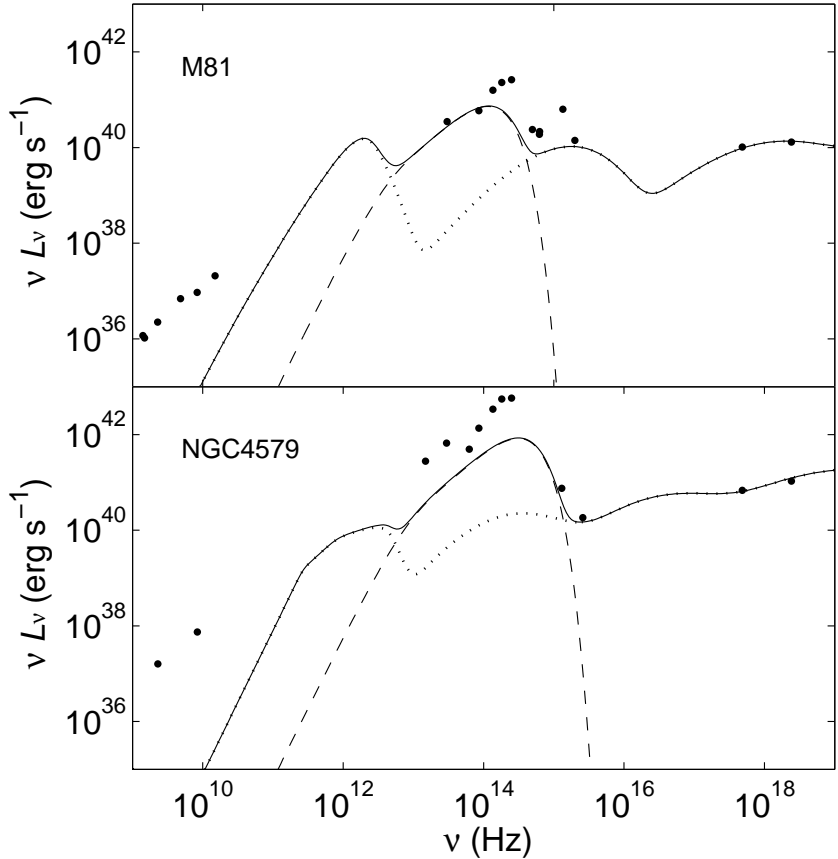} \caption{The ADAF$+$standard thin disk spectral
models for M81 and NGC4579 (solid lines). The dotted lines represent
the spectra of ADAFs, while the dashed lines are the spectra of the
outer standard thin disks. The observed data points are taken from
\citet{h99}.} \label{fig1}
\end{figure}

We find that the best fits to the SEDs of these sources require:
$\dot{m}=3\times10^{-4}$ and $r_{\rm tr}=120$ for M81;
$\dot{m}=3.3\times10^{-3}$ and $r_{\rm tr}=80$ for NGC4579. We find
that the observations in optical/UV and X-ray bands of these two
sources can be fitted quite well (see Fig. \ref{fig1}).


\section{Transition zone of the thin disk to the ADAF}

The global structure of the ADAF$+$standard thin disk systems based
on the assumption of additional turbulence viscosity in the
transition zone was derived by \citet{h96}. \citet{mknn2000}
performed numerical calculations on the ADAF$+$thin disk system
based on \citet{h96}'s model, and obtained the global structure of
the ADAF smoothly connecting to the outer thin disk. They extended
the Honma's analytical solution to be able to describe the global
structure of the ADAF$+$thin disk system including the transition
zone quite well compared with the numerical results \citep[see][for
more details]{mknn2000}. The Honma's analytical solution for the
ADAF$+$standard thin disk system is only valid for fully advection
dominated flows ($f=1$), where $f$ is the ratio of the advected
energy to the viscously dissipated energy in the ADAF. By
introducing a new parameter $a$, the extended Honma's analytical
solution derived by \citet{mknn2000} can deal with the ADAFs with
partial cooling ($f<1$),  \bd v=-\alpha\frac{(3-a)}{5}[1-(R/R_{\rm
tr})^{a}]v_{\rm K}(R),\ed \bd \Omega=\sqrt{\frac{a+5}{5}}(R/R_{\rm
tr})^{a/2}\Omega_{\rm K}(R), \ed \begin{equation} c_{\rm
s}=\sqrt{\frac{2}{5}}[1-(R/R_{\rm tr})^{a}]^{1/2}v_{\rm
K}(R).\label{eq1}\end{equation} The relation between $f$ and $a$ is
\begin{equation} f=\frac{12a(3-a)-24(\alpha_{\rm
T}/\alpha)(1-a)a}{(a+5)(3-a)^{2}}.\label{eq2}\end{equation} As we
are focusing on the transition zone, $\alpha_{\rm T}=0$ is adopted,
because the structure of the transition zone is almost independent
of the value of this parameter (corresponding to the turbulent
energy transport)\citep{mknn2000}. Their derived global structure
(either the analytical one or that derived numerically) is one
temperature (i.e., the temperature of electrons is the same as that
of ions), which prevents us from calculating the spectrum of the
accretion flow directly. {In this work, we include all the radiation
processes in our calculations for the global structure of the inner
ADAF, while we only use \citet{mknn2000}'s model to connect our ADAF
solution to the outer thin disk.} We assume the efficiency of the
released gravitational energy $\eta=0.1$, and a fraction $f$ of it
is advected in the flow. We have
\begin{equation} L_{\rm ADAF}=\eta\dot{m}(1-f)\dot{M}_{\rm Edd}M_{\rm
bh}c^{2},\label{eq3}\end{equation} which can be derived by
integrating their observed continuum emission (subtracting the
optical/UV bumps from the outer thin disk regions). {Thus, the value
of $f$ can be estimated if the accretion rate $\dot{m}$ is known.
The value of $f$ can also be derived from our calculation on the
global structure of the inner ADAF, because the radiation processes
have been considered in the global solution of the ADAF. We find
that the value of $f$ estimated with Eq. (\ref{eq3}) is consistent
with that derived from the global solution of the ADAF. In order to
explore how the thermal line emission varies with the accretion rate
$\dot{m}$, we loose the constraints on the accretion rates
$\dot{m}$, i.e., allowing $\dot{m}$ to vary about an order of
magnitude from the values derived from the model fits on the SEDs of
these two sources, in our calculation of the thermal line emission
from the transition zone.} We can estimate $f$ as a function of the
accretion rate $\dot{m}$, provided $L_{\rm ADAF}$ is derived from
the observed SEDs. {The value of $a$ can be calculated with Eq.
(\ref{eq2}) after $f$ is known. The structure, and then the thermal
line emission, of the transition zone can be calculated as a
function of the accretion rate $\dot{m}$.}
Using Eq. (\ref{eq1}) and (\ref{eq2}), the structure of the
transition zone is available for calculating its thermal X-ray line
emission as a function of accretion rate.


\section{Thermal X-ray line emission from the transition zones}

Besides the narrow iron K$\alpha$ lines at 6.4 KeV, the broad line
components centered at $E_{{\rm K}\alpha}=6.79$~KeV with equivalent
width EW=287 eV for NGC4579, and EW=101 eV for M81, were observed
with XMM-Newton by \citet{dgds04}. The widths of these two lines
with Gaussian fittings are $\sigma=231$~eV and 188~eV for NGC4579
and M81, respectively.

The thermal X-ray line emission can be calculated, when the physical
properties of the plasma, i.e., the temperature, density and
metallicity are specified. Using the models of the ADAF$+$thin
accretion disk systems described in the previous section, we can
calculate their thermal X-ray line emission from the accretion flows
surrounding the black holes in these two low-luminosity active
galactic nuclei, NGC 4579 and M81. In the ADAF$+$standard thin disk
system, the inner ADAF is very hot ($T_{e}\sim10^{9-10}$K) and the
outer thin disk is cold ($T_{e}\sim10^{4-5}$K). Thus, the inner ADAF
is too hot, the plasma is almost completely ionized, to produce
thermal X-ray line emission, while the outer thin disk is too cold
for thermal X-ray line emission. For our interested H-like and
He-like iron line emission, they are most probably emitted from the
transition zone of the ADAF to the outer thin disk, because the
temperature of the transition zone is between
$\sim10^{4-5}-10^{9-10}$~K.

The line luminosity $L_{\rm line}$ can be calculated by integrating
over the transition zone,
\begin{equation} L_{\rm line}=\int_{r_{\rm tr}-\delta r}^{r_{\rm tr}}
 n_{e}(r)^{2}\epsilon_{\rm line}(r)\frac{1-e^{-\tau_{\rm
line}(r)}}{\tau_{\rm line}(r)} 4\pi rH(r)R_{\rm S}^{2}{\rm d}r,
\label{eq7}\end{equation} where $n_{e}$ is the electron number
density, $\delta r$ is the width of the transition zone connecting
the inner ADAF to the thin disk, $r_{\rm tr}$ is the inner radius of
the thin disk, { $H=c_{s}/\Omega$ is the vertical half-thickness of
the transition zone, and ${\tau_{\rm line}=2Hk_{\rm line}}$ is the
optical depth of the emission line in the vertical direction
($k_{\rm line}$ is the line absorption coefficient).} The line
emissivity $\epsilon_{\rm line}(r)$ as a function of temperature is
calculated with the standard software package Astrophysical Plasma
Emission Code (APEC)\citep{s01}. The APEC code includes collisional
excitation, recombination to excited levels and dielectronic
satellite lines\citep*[see][for the details]{s01}. It ignores
photo-ionization, which is not important in these two low-luminosity
AGNs \citep*[see discussion in][]{dgds04}. We assume ionization
equilibrium in the plasma.  In the transition zone, the
density/temperature of the the flow changes by orders of magnitude
from the inner optically-thin ADAF to the outer optically-thick thin
disk. The optical depth of the flow can be very large in the outer
part of the transition region, and the absorption and radiative
transfer of the line is considered in our calculation. The X-ray
continuum emission of the ADAF$+$thin disk system is dominated by
the bremsstrahlung and the Comptonization of the soft photons in the
ADAF. Considering energy resolution of the observations on the X-ray
line emission is quite low, we only calculate the total equivalent
widths of H-like and He-like iron lines.


\begin{figure}
\plotone{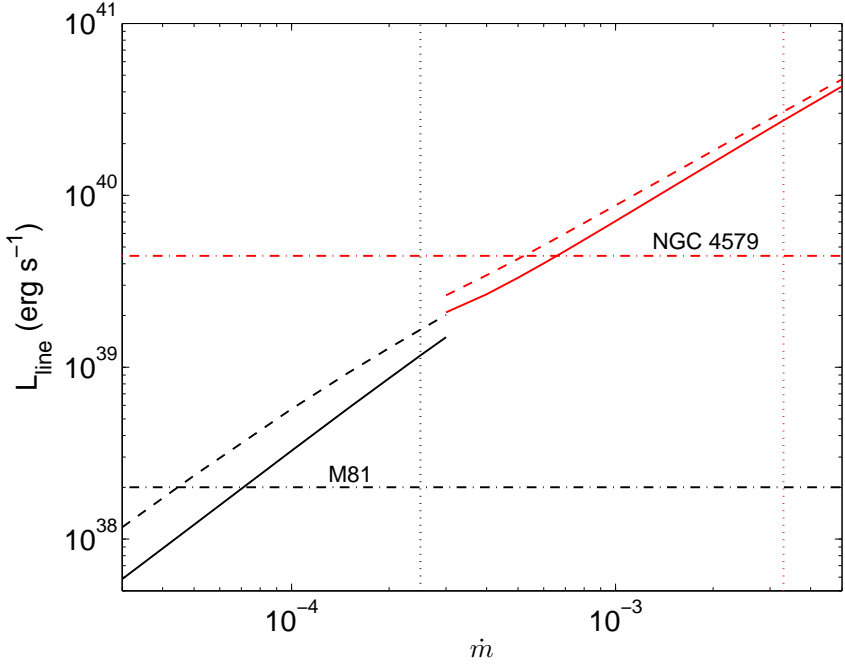} \caption{The total  H-like and He-like iron line
luminosities as functions of accretion rate $\dot{m}$ calculated for
the \citet{h96}'s model. The solid lines correspond to the cases
with the solar metallicity, while the dashed lines are for the cases
with five times solar metallicity. The observed values for these two
sources are marked in the figure (dash-dotted lines). The dotted
lines represent the accretion rates derived from the ADAF+SD model
fitting on the observed SEDs.} \label{fig3}
\end{figure}

In Fig. \ref{fig3},  we plot the thermal X-ray line luminosity as
functions of accretion rate $\dot{m}$ based on the transition model
given by \citet{mknn2000}. Besides the solar metallicity, we also
calculate the X-ray line emission from these two sources for five
times solar metallicity for comparison.

\section{Discussion}

Our best fits to the observed continuum spectra of these two sources
show that the outer thin discs are truncated at $r_{\rm tr}=120$ and
80, for M81 and NGC4579, respectively. The two important disk
parameters, $\dot{m}$ and $r_{\rm tr}$ are mainly determined from
the comparison of the outer disk spectra with the observed
optical/UV continuum emission in these two sources, and the observed
X-ray continuum spectra can then be naturally reproduced by the ADAF
models.
The observed widths of the broad lines are roughly consistent with
the Doppler broadening by the Keplerian motion of the gas in the
transition zone at ~50-150~$R_{\rm S}$ \citep{dgds04}. The ratio of
the transition radii of these two sources is $1.5$, which is
consistent with the observed line width ratio $\simeq 0.814$,
because the width $\sigma\propto r_{\rm tr}^{-1/2}$.

Comparison of our calculations based on the transition zones given
by \citet{mknn2000} (see Fig. \ref{fig3}, and \S 3 and 4 for the
detailed description) with the observations show that the accretion
rates $\dot{m}\sim 7.1\times10^{-5}$ and $6.5\times10^{-4}$ for M81
and NGC4579, respectively (see Fig. \ref{fig3}). These seem to be
inconsistent with the optical/UV spectra observed in these two
sources, which requires the accretion rates to be
$\dot{m}=2.5\times10^{-4}$ (M81) and $3.3\times10^{-3}$ (NGC4579),
unless the transition radii deviate significantly from $\sim 100$
(e.g., an order of magnitude smaller than 100). However, the widths
of the thermal X-ray lines provide strict constraints on the
transition radii, if the thermal X-ray line emission does originate
from the transition zones. The dependence of our results on the
metallicity can be understood with Eq.  (\ref{eq7}). A larger
metallicity will increase the line emissivity and line optical depth
at the same time, while the line luminosity and EW will slightly
increase with a factor ${1-e^{-\tau_{\rm line}}}$. Of course, the
structure of ADAF$+$thin disk systems based on this model is rather
simplified, in which only one temperature is considered (i.e., the
electrons have the same temperature as the ions). Thus, we cannot
rule out this model only from their thermal X-ray line emission. In
the transition zone, the density is very high so that one
temperature flow is a good approximation, as the Coulomb interaction
between ions and electrons in this region is very efficient. Our
calculations of the thermal X-ray line emission from the transition
zone have not been affected by this one temperature assumption. The
more detailed calculations on the ADAF$+$thin disk systems including
energy equilibrium between electrons and ions in the accretion flows
may help to test this model, which is beyond the scope of this work.

An alternative model for the transition of a thin disk to an ADAF is
the ``evaporation" mechanism initially suggested by \citet{mmh94},
in which the thin disk is evaporated as heated by electron
conduction from a hot corona, and then a quasi-spherical hot
accretion flow is formed. There is a very thin layer between the
cold thin disk and the hot corona, of which the electron temperature
may vary from $\sim 10^{6.5}$~K to $\sim 10^9$~K in the corona
\citep*[e.g.][]{liu95,liu97,liu02}. Such a layer, of course, will
emit thermal X-ray line emission. In principle, the evaporation
model for the transition of a thin disk to an ADAF can be tested by
the observed thermal X-ray line emission. However, the structure of
this disk corona system is complicated and is only available by
numerically integrating a set of ordinary differential equations
\citep*[e.g.,][]{liu02}, which prevents us from testing this model
in this paper.

\begin{acknowledgements}
We thank the anonymous referee for his/her helpful
comments/suggestions. We are grateful to R. Narayan and B.F. Liu for
helpful discussion. This work is supported by the NSFC (grants
10778621, 10703003 and 10773020), and the CAS (grant KJCX2-YW-T03).
\end{acknowledgements}

\label{lastpage}



\end{document}